\documentclass[conference]{IEEEtran}
\IEEEoverridecommandlockouts
\usepackage{cite}
\usepackage{amsmath,amssymb,amsfonts}
\usepackage{algorithmic}
\usepackage{graphicx}
\usepackage{textcomp}
\usepackage{xcolor}
\usepackage{multirow}
\usepackage{tabularx}
\usepackage{caption}

\def\BibTeX{{\rm B\kern-.05em{\sc i\kern-.025em b}\kern-.08em
    T\kern-.1667em\lower.7ex\hbox{E}\kern-.125emX}}
\begin{document}

\title{Creating an Intelligent Dementia-Friendly Living Space: A Feasibility Study Integrating Assistive Robotics, Wearable Sensors, and Spatial Technology
}


\author{\IEEEauthorblockN{1\textsuperscript{st} Arshia A Khan}
\IEEEauthorblockA{\textit{Department of Computer Science} \\
\textit{University of Minnesota Duluth}\\
Duluth, MN, USA \\
akhan@d.umn.edu \\
ORCID: 0000-0001-8779-9617}
\and
\IEEEauthorblockN{2\textsuperscript{nd} Rupak Kumar Das}
\IEEEauthorblockA{\textit{College of Information Sciences and Technology} \\
\textit{Pennsylvania State University-University Park}\\
State College, PA, USA \\
rjd6099@psu.edu}
\and
\IEEEauthorblockN{3\textsuperscript{rd} Anna Martin}
\IEEEauthorblockA{\textit{Department of Computer Science} \\
\textit{University of Minnesota}\\
Minneapolis, MN, USA \\
mart5877@d.umn.edu}
\and
\IEEEauthorblockN{4\textsuperscript{th} Dale Dowling}
\IEEEauthorblockA{\textit{Department of Computer Science} \\
\textit{University of Minnesota Duluth}\\
Duluth, USA \\
dowli026@d.umn.edu}
 \and
\IEEEauthorblockN{5\textsuperscript{th} Rana Imtiaz } \IEEEauthorblockA{\textit{Department of Medicine} \\
 \textit{Kentucky School of Osteopathic Medicine}\\
 Pikeville, USA \\
 rana.z.imtiaz@gmail.com}
}

\maketitle

\begin{abstract}
This study investigates the integration of assistive therapeutic robotics, wearable sensors, and spatial sensors within an intelligent environment tailored for dementia care. The feasibility study aims to assess the collective impact of these technologies in enhancing care giving by seamlessly integrating supportive technology in the background. The wearable sensors track physiological data, while spatial sensors monitor geo-spatial information, integrated into a system supporting residents without necessitating technical expertise. The designed space fosters various activities, including robot interactions, medication delivery, physical exercises like walking on a treadmill (Bruce protocol), entertainment, and household tasks, promoting cognitive stimulation through puzzles. Physiological data revealed significant participant engagement during robot interactions, indicating the potential effectiveness of robot-assisted activities in enhancing the quality of life for residents.
\end{abstract}

\begin{IEEEkeywords}
Humanoid Robot, Pepper robot, dementia friendly living space, Alzheimer's care giving, Electrodermal activity, EDA; Physiological data;
\end{IEEEkeywords}

\section{Introduction}
In response to the pressing challenges faced by dementia care, this study delves into the fusion of cutting-edge technologies within a specialized environment designed for dementia patients. The proposed framework aims to create a supportive atmosphere conducive to dementia care by integrating robotic assistance, an automated medication management system, and wearable sensors in a sensor-equipped space designed for dementia-friendly interactions.
Dementia is a growing concern, currently affecting 5.4 million individuals in the US, and is projected to escalate to 106.2 million by 2050. This surge places a substantial strain on caregiving and imposes financial hardships, as evidenced by the \$221.3 billion in unpaid care and \$236 billion in paid care provided in 2015, expected to exceed a trillion dollars by 2050. Women are disproportionately impacted, constituting two-thirds of those affected by dementia and caregivers. With a twofold lifetime risk of Alzheimer’s compared to men, women face higher risks of abandonment, nursing home placement, and decreased informal caregiving resources. To address these challenges and ensure sustained assistance while preserving safety and quality of life, we are developing a sensor-based robotic assistant. By integrating wearable and spatial sensors with therapeutic robots, we aim to offer a human-like social interface specifically tailored for individuals, particularly women, affected by dementia.
The primary objective of this feasibility study is to evaluate the cumulative impact of these integrated technologies, envisioning a paradigm where technology seamlessly bolsters caregiving without requiring technical adeptness from its beneficiaries.
Wearable sensors meticulously capture and monitor participants' physiological data, while spatial sensors intricately observe geospatial information. These data points are amalgamated into a unified system that operates discreetly in the background, offering substantial support to residents without imposing any technical burden. This crafted intelligent living space fosters a rich array of activities, including robot interactions, medication delivery by the robot, physical exercises like the treadmill-based Bruce protocol, entertainment such as video media, and everyday household tasks such as preparing soup, tea, sweeping floors, and folding towels, ensuring a stimulating environment that promotes cognitive engagement through puzzles and other interactive exercises.
Preliminary analysis of physiological data underscores a notable trend—participant engagement peaks significantly during interactions with the robot. This finding indicates the potential efficacy of robot-assisted activities in augmenting the overall quality of life for residents living with dementia.

\section{Methods}

\subsection{Preliminary Work}

We have programmed Pepper, a prebuilt robot, for independent verbal communication to aid medication adherence and have conducted two published studies. To validate our assistive robot integrated with a sensor-based system, we require a specifically designed living space optimized for sensor functionality.
In preparation for testing our system, we have crafted a secure space furnished with sensors tailored to accommodate the robot. Our innovative, dementia-friendly living environment allows us to conduct comprehensive studies, assessing sensor placement, mechanical and technical requirements for the robot's mobility and ensuring a safe dwelling for individuals affected by dementia. Utilizing devices such as Empatica E4, Qualcomm Dragonboard 410c SoC, Magnetic Proximity Sensors, Passive Infrared Sensors, Pressure Sensors, Visible Light Sensors, Omron Blood Pressure Monitors, and AliveCor EKG sensors, we aim to create an integrated system meeting the needs of those in this environment.

\subsection{Study Design}
This study was conducted during the COVID period involving 32 healthy participants. Over the 90-minute study session, participants completed various activities after reading and signing the consent form, undergoing health pre-screening, and filling out pre and post-experiment surveys. The activities comprised an initial sensor setup and readings, administration of a placebo sugar pill as medication simulation, treadmill exercises following the Bruce submaximal treadmill exercise test, additional sensor readings, engagement in sitting/puzzle dementia-friendly activities, door opening/closing to test door sensors, media presentation to induce higher arousal state, further sensor readings, survey activities to determine sensor efficacy during daily tasks, a brief interaction with Pepper the robot for joke-telling, and final sensor readings. These activities aimed to assess sensor functionality, simulate real-world scenarios, and monitor physiological responses during various tasks, contributing to the evaluation of the integrated sensor system.

\begin{figure}
    \centering
    \includegraphics[width=0.25\linewidth]{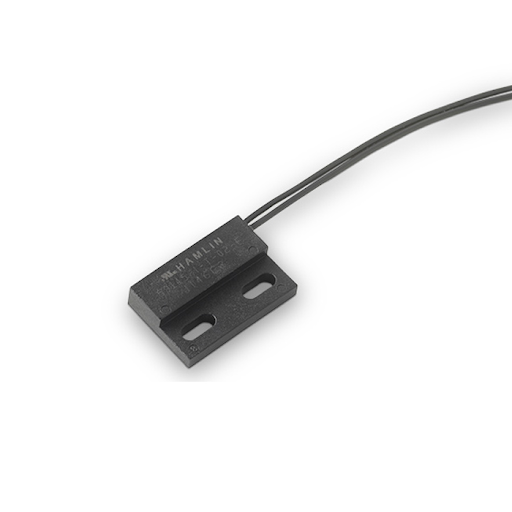}
    \caption{Magnetic flange mount reed sensor }
    \label{fig:Magnetic_flange}
\end{figure}

\begin{figure}
    \centering
    \includegraphics[width=0.25\linewidth]{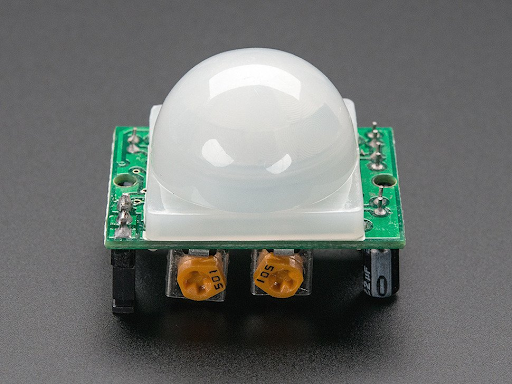}
    \caption{Passive Infrared (PIR) sensor}
    \label{fig:PIR}
\end{figure}

\begin{figure}
    \centering
    \includegraphics[width=0.5\linewidth]{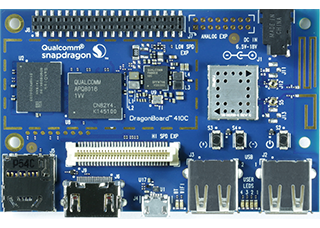}
    \caption{Qualcomm Dragonboard 410c SoC device}
    \label{fig:Dragonboard}
\end{figure}

In addition to the Pepper humanoid robot, this study utilized wearable sensors like the Empatica E4, Omron blood pressure, and the AliveCor EKG sensor, alongside spatial sensors—magnetic proximity, passive infrared, pressure, and spectral light sensors, integrated through a Dragonboard for data collection. The magnetic flange mount reed sensor, specifically the 59145 Series by littleFuse (Fig \ref{fig:Magnetic_flange}), detected door opening/closing states based on proximity to a magnet, aiding in monitoring door activities.
Passive Infrared (PIR) sensors tracked motion by detecting infrared radiation-induced voltage changes in a pyroelectric crystal, facilitating location tracking without camera surveillance as seen in Fig \ref{fig:PIR}. The Interlink Electronics FSR X 400 Series 406 pressure sensor functioned as a binary indicator of participant seating through voltage changes based on applied force.
The Qualcomm Dragonboard 410c SoC device (seen in Fig \ref{fig:Dragonboard}), akin to Arduino and Raspberry Pi boards, supported embedded programming and IoT applications, running Linaro, a Linux version. The sensors are connected to Dragonboards via the 'Sensors Mezzanine' I/O expansion headers with an integrated ATMEGA328p microcontroller for Arduino compatibility and analog-to-digital conversion. These boards initiated data collection upon power-up, aligned with study sessions.
\\

\section{Data Collection}
A database was created to hold the data from all the sensors. Data from the dragonboards, EDA, Omron Blood pressure monitor, AliveCor EKG sensor, the robot, infrared sensors, visible light sensors, door sensors, and the chair and bed pressure sensors were stored in this database. 

Participants’ subjective experiences during the study were evaluated through two surveys: the Post-Study Survey, given in paper form after the study's completion, and the Presentation Survey conducted during the IADS presentation. The Post-Study Survey gauged valence, arousal, and sense of control during pill administration, treadmill exercises, IADS presentation, and Pepper interview activities, using a scale of 1 to 5 for each metric. Valence assessed happiness (1 being very happy, 5 very unhappy), arousal measured excitement or calmness (1 very excited, 5 very calm), and sense of control examined feelings of control (1 most controlled, 5 most in control).
The Presentation Survey, administered during the IADS presentation, assessed valence and arousal for each image or sound, using a scale of 1 to 9. Valence measured pleasantness (1 most pleasant, 9 most unpleasant), while arousal assessed calmness or excitement (1 most excited, 9 most calm).
\begin{figure}
    \centering
    \includegraphics[width=0.25\linewidth]{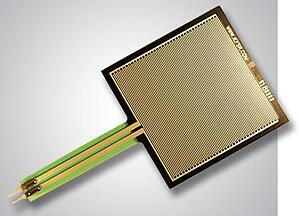}
    \caption{Pressure sensor}
    \label{fig:Pressure_sensor}
\end{figure}

\section{Data Analysis}

\subsection{EDA Analysis Method}\label{EDA_analysis}

Electrodermal Activity (EDA) data was sampled continuously at a rate of four Hz throughout the study. Using timestamps denoting the start and end times for each section of the study, the EDA signal for each participant was segmented into 15 consecutive chunks. These chunks corresponded to the various activities involved in the procedure, including the 3-minute collection of baseline EDA data, four additional data collection sessions, four stages of the Bruce protocol gathered during the treadmill activity, the wandering and opening doors activity, pill administration, puzzle activities, household task activities, the International Affective Picture System (IAPS) \cite{b1} and International Affective Digital Sound (IADS) presentation, and the Pepper interview activity. Due to issues with the streaming server, occasionally, readings would not be recorded. These missing records were detected using the timestamps of the surrounding samples, and gaps of two seconds or fewer were automatically filled in with cubic spline interpolation. The segmented raw EDA data was stored in individual files, and further processing was performed using the Matlab plugin Ledalab\footnote{http://www.ledalab.de/}. The records for each activity for each participant were visually inspected using Ledalab’s GUI, and portions of signals where more than two seconds were lost were removed. After cleaning out lost signals, we processed the dataset using Ledalab’s batch mode. The following preprocessing steps were performed during this analysis:

\begin{itemize}
  \item A first-order Butterworth filter was applied with a lower-cutoff frequency of 0.35
  \item The signal was decomposed into tonic and phasic signals, and Skin Conductance Responses (SCRs) were automatically extracted using Continuous Decomposition Analysis (CDA).
\end{itemize}

The data output by Ledalab that we chose for our analysis was the raw tonic signal, the SCR amplitudes, and the frequency of SCR peaks per activity. Each of these features was explored at the participant level as well as across participants. The mean and standard deviation were calculated for the entire signal for each participant, and the mean of each activity’s segmented signal was calculated. The raw tonic signal and SCR amplitudes were then standardized by finding the z-score of each value in the signal. 

\begin{equation}
z = \frac{{x - \mu}}{{\sigma}}
\end{equation}

Where each value $x$ in the signal is transformed by subtracting the mean of the total signal and dividing the difference by the standard deviation ${\sigma}$.

The SCR frequencies for each activity were standardized by subtracting the rate of SCRs per second for the duration of the study from the rate of SCRs for each activity. This preprocessing step was required in order to compare across participants. 

The Pearson Correlation Coefficient was used to calculate the correlation between each participant, and Fisher’s z-score was used to determine the confidence intervals. The Fisher $z$ transformation was also used to normalize the correlation coefficients before finding the mean correlation coefficient of every participant. 

The frequency of SCR peaks per second was calculated per activity for each participant by dividing the number of samples in the associated raw tonic signal by the sample rate to determine the number of seconds per activity for each participant, then dividing the number of SCR peak amplitudes by the resulting quotient. The mean SCR amplitude frequency was calculated for each activity per participant.

\subsection{Bruce Protocol}

The Bruce protocol treadmill test \cite{b2} is normally used to estimate the overall fitness level of athletes. The test was designed by a famous cardiologist named Robert A. Bruce in 1963. This is a non-invasive test to assess patients who have suspected heart diseases. In this test, the participant needs to walk on a treadmill while different physiological signals are monitored and recorded using different sensors. Ventilation volumes, Electrodermal activities, Electrocardiogram signals, respiratory gas exchanges, etc., are also monitored before, during, and after exercise. As the speed of the treadmill is adjustable, this physical activity can be tolerated by most patients. Here is the actual Bruce Protocol Table available for Sub Maximal (more practical with the majority of the non-athletic or competitively athletic population) efforts. We used a modified Bruce protocol for our experiment. Our modified protocol is shown in the table \ref{bruce_protocol}. It has 4 stages, and the duration of every stage is 3 minutes.


\begin{table}
\caption{Modified Bruce protocol for this experiment}
\centering
\label{bruce_protocol}
\begin{tabular}{ccc}
\hline
Stage & Minutes & m/s\\
\hline
1 & 3 & 0.4 \\
2 & 3 & 0.8 \\
3 & 3 & 1.2 \\
4 & 3 & 1.6 \\
\hline
\end{tabular}
\end{table}

Using the Bruce protocol, the Center of Pressure (COP) and different COP features were extracted. The loading system acting on a force plate can be described by three forces ($Fx$, $Fy$, $Fz$) and three moments ($Mx$, $My$, $Mz$) components. The point of application of force or the center of pressure ($Xp$, $Yp$) and the couple (or torque) can be calculated from those components.

The X coordinate of the center of pressure can be calculated using equation \ref{x_cor_cop}.

\begin{equation} \label{x_cor_cop}
X_p = \frac{{-M_y + (F_x \cdot dz)}}{{F_z}}
\end{equation}

The Y coordinate of the center of pressure can be calculated using equation \ref{y_cor_cop}.

\begin{equation} \label{y_cor_cop}
Y_p = \frac{{M_x + (F_y \cdot dz)}}{{F_z}}
\end{equation}

Finally, the couple can be calculated using equation \ref{couple_cop}

\begin{equation} \label{couple_cop}
T_z = M_z - X_p F_y + Y_p F_x
\end{equation}

Where $F_x$, $F_y$, and $F_z$ represent the components of the ground reaction force in different directions, respectively. $M_x$, $M_y,$ and $M_z$  are components of the moment of force. $d_z$ is the depth of the force plate.

\subsection{Blood Pressure and Heart Rate}

Participants blood pressure and heart rate were measured four times during the course of the study. The baseline was gathered in the beginning, reading 2 was gathered after the treadmill activity, reading 3 was gathered after the cooking and cleaning activities, and reading 4 was gathered after the IADS presentation.

\subsection{Participant Feedback Surveys}

Participants’ subjective experience of the various study activities was measured using two surveys. The Post Study Survey was administered as a paper form after the end of the study. It measured participants’ valence, arousal, and sense of control during pill administration, treadmill, IADS presentation, and pepper interview activities. Each of these metrics was measured on a scale from 1 to 5. Valence was measured by asking participants how happy they felt during each activity, with 1 being very happy and 5 being very unhappy. Arousal was measured by asking participants how excited or calm they felt during each activity, with 1 being very excited and 5 being very calm. Sense of control was measured by asking participants how controlled or in control they felt, where 1 is feeling the most controlled and 5 is feeling the most in control. 

The Presentation Survey was administered during the IADS presentation and measured participants’ valence and arousal for each image or sound in the presentation. Both of these metrics were measured on a scale from 1 to 9. Valence was measured by asking participants how pleasant or unpleasant the image or sound was, where 1 is the most pleasant and 9 is the calmest. Arousal was measured by asking participants how calm or excited the image or sound made them feel, where 1 was most excited, and 9 was most calm. 

\section{Results and Analysis}\label{results}

\subsection{EDA Results During the Activities}
Statistical analysis shows some relationship between EDA features and the activities being performed, although the overall correlation between participants is weak. 

We found the mean for each activity per participant for the tonic signal and SCR amplitude, extracted the SCR frequency, and looked at which activities co-occurred most frequently with participants’ maximum and minimum values. We found that the maximum tonic mean occurred most often in nine participants during the Pepper interview activity. The puzzles activity was second most, as seven participants recorded their highest mean tonic value while doing the puzzles. The third most frequent occurred during the second data-gathering session, and the fourth was the third data-gathering session. The minimum tonic mean was recorded most often during the baseline reading and the first data collection, which were tied at 11 participants each.

\begin{figure*}[h!]  
  \centering
  \includegraphics[width=\textwidth, height=0.2\textheight]{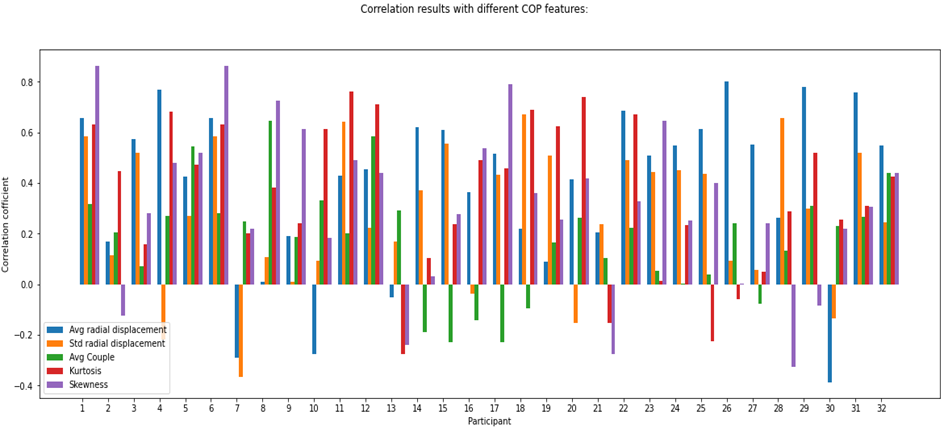}  
  \caption{Correlation results with different COP features of force plate 1 (left force plate)}
  \label{fig:left_force_plate}
\end{figure*}

\begin{figure*}[h!]  
  \centering
  \includegraphics[width=\textwidth,height=0.2\textheight]{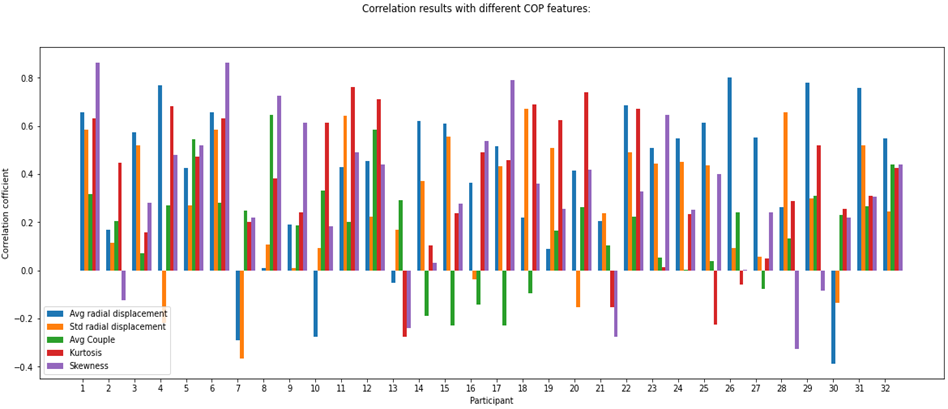}  
  \caption{Correlation results with different COP features of force plate 2 (right force plate)}
  \label{fig:right_force_plate}
\end{figure*}

The puzzle activity was the study event during which the most participants recorded their highest SRC amplitudes, at 7 participants, followed by the cooking and housework activities at 6 participants and the Pepper interview activity at 5 participants. The most common activity associated with minimum mean SRC amplitude was the baseline at 18 participants, followed by the second phase of the Bruce protocol at 6 participants. 

The wandering activity was associated with the highest SCR frequency for most participants, at 6 participants. The second and third Bruce protocol stages were the next most co-occurring activities with the highest SCR frequency, at 4 and 5 participants, respectively. Data gathering sessions 3 and 4 were associated with the minimum recorded SCR frequency for the most participants, at 5 and 7 participants, respectively.

A subset of participants had EDA features that were significantly correlated between them; the pairs of participants with Pearson’s correlation coefficients greater than 0.75, p-scores less than 0.05, and confidence intervals less than 0.4 are reported in Table \ref{table:Pairwise_Pearson_correlation_coefficients}.  However, averaging the Fisher z-score transformations of every correlation coefficient between each participant showed a weak correlation for the aggregate participants, with z-scores of 0.49 for the tonic signal, 0.46 for the SCR amplitudes, and 0.23 for the frequency of amplitudes.

\begin{table}
\caption{Pairwise Pearson’s correlation coefficients of EDA features}
\centering
\label{table:Pairwise_Pearson_correlation_coefficients}
\begin{tabular}{cccc}
\hline
Participant pair & Correlation coefficient & p-score & Confidence interval \\
\hline
\multicolumn{4}{c}{Correlation of mean tonic signals} \\
\hline
10, 18 & 0.94 & \textless .001 & 95\% CI[.83, .98] \\
10, 20 & 0.95 & \textless .001 & 95\% CI[.85, .98] \\
10, 26 & 0.92 & \textless .001 & 95\% CI[.77, .97] \\
10, 27 & 0.89 & \textless .001 & 95\% CI[.69, .96] \\
15, 18 & 0.84 & \textless .001 & 95\% CI[.56, 94] \\
15, 24 & 0.84 & \textless .001 & 95\% CI[.57, .94] \\
16, 23 & 0.88 & \textless .001 & 95\% CI[.66, .96] \\
16, 27 & 0.84 & \textless .001 & 95\% CI[.66, .96] \\
16, 28 & 0.89 & \textless .001 & 95\% CI[.71, .96] \\
16, 30 & 0.91 & \textless .001 & 95\% CI [.74, .97] \\
18, 20 & 0.89 & \textless .001 & 95\% CI[.69, .96] \\
18, 26 & 0.84 & \textless .001 & 95\% CI [.59, .95] \\
18, 27 & 0.96 & \textless .001 & 95\% CI[.87, .99] \\
20, 26 & 0.85 & \textless .001 & 95\% CI[.61, .95] \\
20, 27 & 0.90 & \textless .001 & 95\% CI[.71, .97] \\
23, 30 & 0.94 & \textless .001 & 95\% CI[.84, .98] \\
27, 28 & 0.92 & \textless .001 & 95\% CI[.77, .97] \\
\hline
\multicolumn{4}{c}{Correlation of SCR amplitudes} \\
\hline
10, 30 & 0.84 & \textless .001 & 95\% CI[.58, .95] \\
13, 15 & 0.83 & \textless .001 & 95\% CI[.55, .94] \\
17, 21 & 0.84 & \textless .001 & 95\% CI[.57, .94] \\
17, 26 & 0.90 & \textless .001 & 95\% CI[.72, .97] \\
17, 28 & 0.86 & \textless .001 & 95\% CI[.61, .95] \\
18, 20 & 0.89 & \textless .001 & 95\% CI[.7, .96] \\
18, 30 & 0.86 & \textless .001 & 95\% CI[.62, .95] \\
21, 26 & 0.85 & \textless .001 & 95\% CI[.61, .95] \\
21, 28 & 0.85 & \textless .001 & 95\% CI[.6, .95] \\
21, 31 & 0.87 & \textless .001 & 95\% CI[.65, .96] \\
\hline
\multicolumn{4}{c}{Correlation of SCR frequency} \\
\hline
12, 22 & 0.85 & \textless .001 & 95\% CI[.61, .95]\\
25, 29 & 0.92 & \textless .001 & 95\% CI[.76, .97] \\
\hline
\end{tabular}
\end{table}



\subsection{Bruce Protocol}

In this section, the Center of Pressure (COP) and different COP features (average radial displacement, standard deviation, kurtosis, skewness of radial displacement, and couple)  are extracted from Force ($Fx$, $Fy$, $Fz$) and Moment ($Mx$, $My$, $Mz$) data for both force plates. The result is summarized in the figure below. The average Person Correlation Coefficients (PCC) were determined for all subjects. To find the average PCC, we used Fisher’s z transformation. We extracted the COP ($x$ and $y$ coordinate), Average radial displacement, Standard deviation of radial displacement, and the couples of all participants for different force plate speeds (0.4, 0.8, 1.2, and 1.6 m/s). Among those features, the average radial displacement and standard deviation of radial displacement were found to perform best compared to the other COP features. That means with the change of speed, the COP changes frequently for most of the participants.  On the other hand, the couple, skewness, and kurtosis showed comparatively lower PCC values than the other two features. Figure \ref{fig:left_force_plate} and figure \ref{fig:right_force_plate} show the correlation results with different COP features for force plate 1 (left force plate) and 2 (right force plate), respectively.

\subsection{Blood Pressure and Heart Rate Results}

The blood pressure and pulse readings gathered from the Omron wristband, and EKG are reported for each participant in Table \ref{tab:blood_pressure_and_pulse} in the appendix \ref{appendix_data}. There were occasional issues with the Omron wristband that prevented us from getting a reading; those instances are recorded as ND for No Data. The mean values for each reading across all participants are recorded in Table \ref{tab:The_mean_values_of_blood_pressure_and_pulse} in the appendix \ref{appendix_data}. Systolic blood pressure and heart rate readings were highest on average during Reading 2, which was gathered immediately after the treadmill activity. Diastolic blood pressure rose gradually throughout the duration of the study. Reading 4 occurred after the IADS presentation. Despite the fact that the participants found the presentation to be unpleasant and alarming, there is no significant difference in the blood pressure and heart rates gathered immediately after viewing the presentation.

\subsection{Post Study Survey Results}

\begin{figure}[h!]  
  \centering
  \includegraphics[width=\columnwidth]{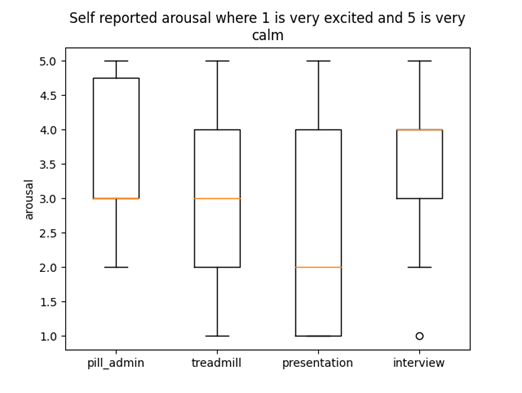}  
  \caption{Self-reported arousal where 1 is very excited and 5 is very calm}
  \label{fig:self_reported_arousal}
\end{figure}

\begin{figure}[h!]  
  \centering
  \includegraphics[width=\columnwidth]{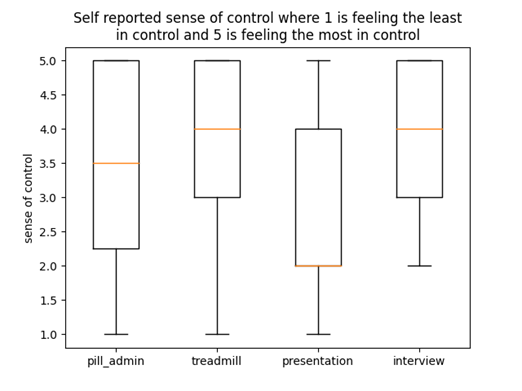}  
  \caption{Self-reported sense of control 1 is feeling the least in control and 5 is feeling the most in control}
  \label{fig:self_reported_sense_control}
\end{figure}

\begin{figure}[h!]  
  \centering
      \includegraphics[width=\columnwidth]{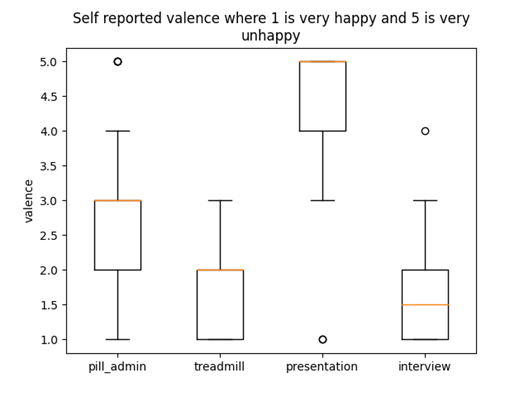}  
  \caption{Self-reported valence where 1 is very happy and 5 is very unhappy}
  \label{fig:self_reported_valence}
\end{figure}

Box plots are provided below to report the mean and standard deviation for valence, arousal, and sense of control. Figure \ref{fig:self_reported_arousal}, \ref{fig:self_reported_sense_control}, and \ref{fig:self_reported_valence} provide the box plots reports for arousal, sense of control and valence, respectively, during the pill administration, treadmill activity, presentation, and interview sessions. Most participants were more excited, least in control, and unhappy during the presentation. They were calm, mostly in control, and happy during the interview session.

\begin{figure}[h!]  
  \centering
  \includegraphics[width=\columnwidth]{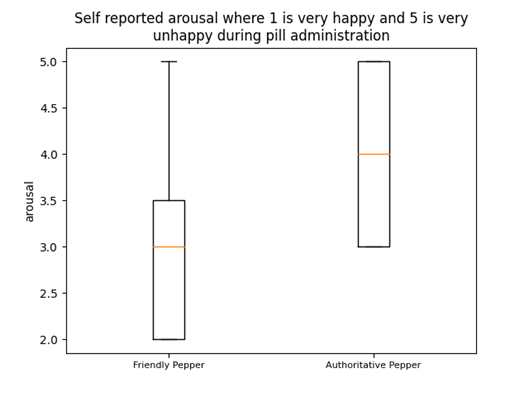}  
  \caption{Self-reported arousal where 1 is very happy and 5 is very unhappy during pill administration}
  \label{fig:self_reported_arousal_2}
\end{figure}

\begin{figure}[h!]  
  \centering
  \includegraphics[width=\columnwidth]{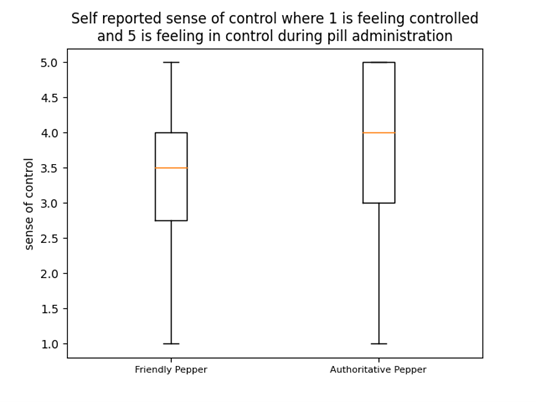}  
  \caption{Self-reported sense of control where 1 is feeling the least in control and 5 is feeling the most in control during pill administration}
  \label{fig:self_reported_sense_control_2}
\end{figure}

\begin{figure}[h!]  
  \centering
      \includegraphics[width=\columnwidth]{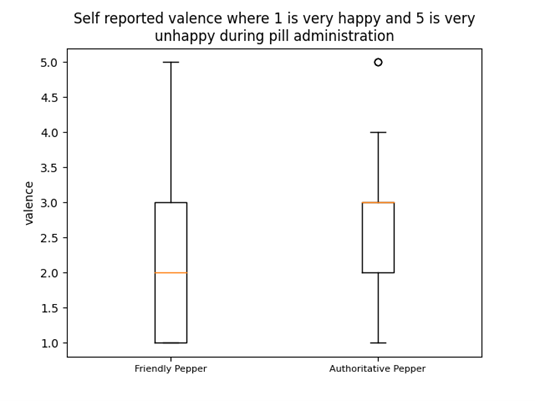}  
  \caption{Self-reported valence where 1 is very happy and 5 is very unhappy during pill administration}
  \label{fig:self_reported_valence_2}
\end{figure}

On the other hand, figures \ref{fig:self_reported_arousal_2}, \ref{fig:self_reported_sense_control_2} and \ref{fig:self_reported_valence_2} show the box plot reports for arousal, sense of control, and valence, respectively, while facing the friendly pepper and authoritative pepper during the pill administration. The results show that participants liked the happy pepper over the authoritative pepper. 

\section{Discussion}
With the use of many different streaming devices during the course of an hour-long study comes the risk of device failure due to low battery, interruption of signal, or user error. The following steps were taken to mitigate these risks: all devices were plugged into power sources at the end of each session to ensure full power during the next session; the Empatica E4 wristband, Omron wristband, and Pepper robot were duplicated so that if one failed during the course of a session, the backup device could quickly replace it; and videos were played for the participants during the study that show how to put on the E4 and Omron wristbands and how to use the AliveCor EKG sensor. 
The Empatica E4 wristband was not difficult for participants to put on and generally did not cause problems during the course of the study. However, gaps were discovered in the recorded Empatica E4 signal that could have been caused by either the streaming signal being dropped or interrupted or by loss of contact with the participant’s skin. 
The Omron wristband was very difficult for participants to put on themselves; this caused visible stress in some of the participants. This could not be addressed by physically helping the participants due to the COVID-19 safety protocol. Furthermore, the size of the wristband was too large for some of the participants, making it difficult to collect good readings. Another issue that arose with Omron was user error during readings. Gathering blood pressure data from Omron requires participants to sit very still. The cuff constricts twice, and the participant must wait until the cuff fully relaxes after the second constriction before they can move. Many participants wanted to check the progress of the wristband by turning their wrists to look at the dial. This issue could be addressed by providing a second Omron instructional video that shows the process of gathering a reading in addition to the provided video that shows how to wear it. Participants must also be instructed not to talk or move other parts of their bodies while gathering the blood pressure data.

Another source of friction during the study was the filling out of various forms and surveys by the participants. The wording in the participant eligibility form was found to be confusing and ambiguous to some participants and should be reviewed and clarified. 
The presentation survey filled out during the IADS presentation was sometimes misunderstood due to several factors. Firstly, ten seconds for each image was not always enough time for them to comprehend and rate their valence and arousal. Secondly, some participants wanted to retroactively change their ratings in cases where they gave an earlier image the maximum rating for unpleasantness and then came across an image that was even more unpleasant later in the presentation. Thirdly, participants were confused about the terms “calm” versus “excited” because the word “excited” can have a positive connotation. Lastly, some participants were confused because a 9 on the valence scale meant “very unpleasant,” but a 9 on the arousal scale meant “calm”; it might reduce friction to invert one of the scales so that viewing an extremely unpleasant and physiologically disturbing image would result in a high number on both the valence and arousal scales. Allowing the participant to practice responding to images before the presentation begins would help mitigate these issues, as would providing an instructional video on how to fill out the presentation survey.
The post-study survey was better understood in general, but a few participants did not provide responses to all three questions for each activity. 

One limitation of implementing the Bruce protocol is that the researchers have not used any Markers for the gait analysis. With the help of markers, some gait parameters like Cadence, stride length, swing time, stance time, velocity, etc could be measured. 

\section{Conclusion}
This study was approved by the ethics review board at the University of Minnesota. In conclusion, the successful design and testing of an intelligent living space tailored for dementia care were conducted in a controlled lab environment with healthy participants. The primary aim was to assess the viability of integrating robots, wearable sensors, and spatial technology to support the well-being of individuals affected by dementia. Utilizing pressure sensors in chairs alongside the robot's emotion detection system effectively evaluated participants' moods, aligning with self-reported moods in 95\% of cases. Additionally, door sensors accurately detected exiting behavior 100\% of the time, while wall-mounted PIR sensors successfully identified wandering behavior. Wearable sensors, including EDA and blood pressure sensors, coupled with robot interactions, detected physiological changes during various activities, indicating the most significant impact during interactions with the robot. These findings suggest that an intelligent living space specifically designed for dementia care can effectively preserve the quality of life for affected individuals, particularly through robot interactions, showcasing promising effectiveness.

\section*{Acknowledgment}

Shawn Savela for helping with the study.

\appendix
\section{Appendix A: Blood Pressure and Pulse data} \label{appendix_data}


Table \ref{tab:blood_pressure_and_pulse} provides the blood pressure and pulse of all the participants, and the mean value of blood pressure and pulse is presented in table \ref{tab:The_mean_values_of_blood_pressure_and_pulse}.

\begin{table}[ht]
\centering
\captionsetup{justification=centering}
\caption{The blood pressure and pulse of the participants}
\label{tab:blood_pressure_and_pulse}
\begin{tabular}{c|cc|cc|cc|cc}
\hline
\multirow{2}{*}{Participant} & \multicolumn{2}{c|}{Baseline}                     & \multicolumn{2}{c|}{Reading 2}                    & \multicolumn{2}{c|}{Reading 3}                    & \multicolumn{2}{c}{Reading 4}                    \\ \cline{2-9} 
                             & \multicolumn{1}{c}{Sys/Dia (mmHg)} & Pulse (BPM) & \multicolumn{1}{c}{Sys/Dia (mmHg)} & Pulse (BPM) & \multicolumn{1}{c}{Sys/Dia (mmHg)} & Pulse (BPM) & \multicolumn{1}{c}{Sys/Dia (mmHg)} & Pulse (BPM) \\ \hline
1                            & \multicolumn{1}{c}{ND}             & 98          & \multicolumn{1}{c}{ND}             & 93          & \multicolumn{1}{c}{ND}             & 75          & \multicolumn{1}{c}{86/69}          & 73          \\ 
2                            & \multicolumn{1}{c}{99/62}          & 94          & \multicolumn{1}{c}{108/64}         & 95          & \multicolumn{1}{c}{ND}             & ND          & \multicolumn{1}{c}{ND}             & 88          \\ 
3                            & \multicolumn{1}{c}{85/55}          & 83          & \multicolumn{1}{c}{86/60}          & 97          & \multicolumn{1}{c}{83/58}          & 83          & \multicolumn{1}{c}{78/54}          & 87          \\ 
4                            & \multicolumn{1}{c}{97/62}          & 89          & \multicolumn{1}{c}{103/60}         & 104         & \multicolumn{1}{c}{ND}             & ND          & \multicolumn{1}{c}{116/79}         & 86          \\ 
5                            & \multicolumn{1}{c}{90/57}          & 91          & \multicolumn{1}{c}{107/64}         & 112         & \multicolumn{1}{c}{104/64}         & 112         & \multicolumn{1}{c}{108/68}         & 106         \\ 
6                            & \multicolumn{1}{c}{115/72}         & 81          & \multicolumn{1}{c}{119/64}         & 99          & \multicolumn{1}{c}{125/64}         & 91          & \multicolumn{1}{c}{113/71}         & 86          \\ 
7                            & \multicolumn{1}{c}{99/65}          & 84          & \multicolumn{1}{c}{97/61}          & 93          & \multicolumn{1}{c}{93/57}          & 87          & \multicolumn{1}{c}{94/60}          & 93          \\ 
8                            & \multicolumn{1}{c}{134/60}         & 89          & \multicolumn{1}{c}{124/59}         & 95          & \multicolumn{1}{c}{95/53}          & 89          & \multicolumn{1}{c}{100/53}         & 88          \\ 
9                            & \multicolumn{1}{c}{103/75}         & 95          & \multicolumn{1}{c}{102/71}         & 101         & \multicolumn{1}{c}{103/76}         & 89          & \multicolumn{1}{c}{99/80}          & 92          \\ 
10                           & \multicolumn{1}{c}{130/82}         & 87          & \multicolumn{1}{c}{111/74}         & 92          & \multicolumn{1}{c}{115/83}         & 92          & \multicolumn{1}{c}{117/85}         & 96          \\ 
11                           & \multicolumn{1}{c}{94/69}          & 84          & \multicolumn{1}{c}{95/63}          & 89          & \multicolumn{1}{c}{100/70}         & 83          & \multicolumn{1}{c}{101/72}         & 78          \\ 
12                           & \multicolumn{1}{c}{148/80}         & 91          & \multicolumn{1}{c}{ND}             & ND          & \multicolumn{1}{c}{154/85}         & 113         & \multicolumn{1}{c}{166/87}         & 102         \\ 
13                           & \multicolumn{1}{c}{108/62}         & 67          & \multicolumn{1}{c}{ND}             & 72          & \multicolumn{1}{c}{ND}             & 68          & \multicolumn{1}{c}{ND}             & 61          \\ 
14                           & \multicolumn{1}{c}{110/75}         & 74          & \multicolumn{1}{c}{110/72}         & 76          & \multicolumn{1}{c}{101/67}         & 76          & \multicolumn{1}{c}{108/75}         & 72          \\ 
15                           & \multicolumn{1}{c}{113/61}         & 90          & \multicolumn{1}{c}{ND}             & ND          & \multicolumn{1}{c}{ND}             & ND          & \multicolumn{1}{c}{ND}             & ND          \\ 
16                           & \multicolumn{1}{c}{ND}             & 79          & \multicolumn{1}{c}{ND}             & ND          & \multicolumn{1}{c}{ND}             & 79          & \multicolumn{1}{c}{ND}             & ND          \\ 
17                           & \multicolumn{1}{c}{120/68}         & 70          & \multicolumn{1}{c}{103/64}         & 67          & \multicolumn{1}{c}{114/76}         & 70          & \multicolumn{1}{c}{104/70}         & 67          \\ 
18                           & \multicolumn{1}{c}{130/65}         & 71          & \multicolumn{1}{c}{140/67}         & 80          & \multicolumn{1}{c}{129/72}         & 80          & \multicolumn{1}{c}{138/74}         & 74          \\ 
19                           & \multicolumn{1}{c}{99/62}          & 72          & \multicolumn{1}{c}{125/69}         & 84          & \multicolumn{1}{c}{102/65}         & 80          & \multicolumn{1}{c}{90/68}          & 76          \\ 
20                           & \multicolumn{1}{c}{100/52}         & 67          & \multicolumn{1}{c}{118/68}         & 72          & \multicolumn{1}{c}{130/77}         & 70          & \multicolumn{1}{c}{118/69}         & 68          \\ 
21                           & \multicolumn{1}{c}{134/68}         & 79          & \multicolumn{1}{c}{146/74}         & 86          & \multicolumn{1}{c}{141/80}         & 77          & \multicolumn{1}{c}{145/83}         & 77          \\ 
22                           & \multicolumn{1}{c}{119/73}         & 86          & \multicolumn{1}{c}{127/85}         & 98          & \multicolumn{1}{c}{123/80}         & 98          & \multicolumn{1}{c}{119/75}         & 92          \\ 
23                           & \multicolumn{1}{c}{100/47}         & 78          & \multicolumn{1}{c}{100/52}         & 75          & \multicolumn{1}{c}{112/60}         & 78          & \multicolumn{1}{c}{105/64}         & 80          \\ 
24                           & \multicolumn{1}{c}{112/70}         & 91          & \multicolumn{1}{c}{115/76}         & 90          & \multicolumn{1}{c}{111/72}         & 86          & \multicolumn{1}{c}{122/79}         & 86          \\ 
25                           & \multicolumn{1}{c}{135/71}         & 67          & \multicolumn{1}{c}{148/79}         & 68          & \multicolumn{1}{c}{146/84}         & 60          & \multicolumn{1}{c}{149/78}         & 65          \\ 
26                           & \multicolumn{1}{c}{104/60}         & 57          & \multicolumn{1}{c}{103/60}         & 56          & \multicolumn{1}{c}{102/63}         & 57          & \multicolumn{1}{c}{100/62}         & 53          \\ 
27                           & \multicolumn{1}{c}{113/69}         & 89          & \multicolumn{1}{c}{128/80}         & 103         & \multicolumn{1}{c}{122/80}         & 109         & \multicolumn{1}{c}{110/77}         & 105         \\ 
28                           & \multicolumn{1}{c}{101/67}         & 82          & \multicolumn{1}{c}{102/61}         & 98          & \multicolumn{1}{c}{99/64}          & 93          & \multicolumn{1}{c}{81/52}          & 85          \\ 
29                           & \multicolumn{1}{c}{117/68}         & 72          & \multicolumn{1}{c}{122/74}         & 76          & \multicolumn{1}{c}{114/71}         & 72          & \multicolumn{1}{c}{147/78}         & 71          \\ 
30                           & \multicolumn{1}{c}{128/78}         & 73          & \multicolumn{1}{c}{115/73}         & 80          & \multicolumn{1}{c}{111/69}         & 75          & \multicolumn{1}{c}{109/72}         & 72          \\ 
31                           & \multicolumn{1}{c}{113/58}         & 69          & \multicolumn{1}{c}{143/82}         & 71          & \multicolumn{1}{c}{152/85}         & 71          & \multicolumn{1}{c}{145/92}         & 67          \\ 
32                           & \multicolumn{1}{c}{ND}             & 61          & \multicolumn{1}{c}{ND}             & 61          & \multicolumn{1}{c}{ND}             & 57          & \multicolumn{1}{c}{ND}             & 57          \\ \hline
\end{tabular}
\end{table}

\begin{table}[htb]
\centering
\captionsetup{justification=centering}
\caption{The mean values of blood pressure and pulse for each reading across all participants}
\label{tab:The_mean_values_of_blood_pressure_and_pulse}
\begin{tabular}{c|cc|cc|cc|cc}
\hline
& \multicolumn{2}{c|}{Baseline} & \multicolumn{2}{c|}{Reading 2} & \multicolumn{2}{c|}{Reading 3} & \multicolumn{2}{c}{Reading 4} \\ \cline{2-9}
& Mean & $\pm$ Std & Mean & $\pm$ Std & Mean & $\pm$ Std & Mean & $\pm$ Std \\ \hline
Systolic (mmHg) & 112.07 & 15.19 & 116.08 & 16.48 & 113.63 & 16.86 & 112.5 & 19.7 \\
Diastolic (mmHg) & 66.08 & 8.25 & 68.83 & 8.15 & 70.5 & 9.06 & 71.29 & 9.89 \\
Heart Rate (BPM) & 78.41 & 10.42 & 84.85 & 14.14 & 79.89 & 13.05 & 78.78 & 13.38 \\ \hline
\end{tabular}
\end{table}


\begin{thebibliography}{00}
\bibitem{b1} Lang, P. J., Bradley, M. M., \& Cuthbert, B. N. (1997). International affective picture system (IAPS): Technical manual and affective ratings. NIMH Center for the Study of Emotion and Attention, 1(39-58), 3.

\bibitem{b2} Bruce RA, Blackmon JR, Jones JW, et al. Exercise testing in adult normal subjects and cardiac patients. Pediatrics 1963;32:742-756.

\end{thebibliography}
\end{document}